\documentclass[%
 aip,
 apl,%
 amsmath,amssymb,
preprint,%
]{revtex4-1}

\usepackage{graphicx}
\usepackage{dcolumn}
\usepackage{bm}

\begin{document}


\title{Formation of collimated sound beams by three-dimensional sonic crystals}

\author{A. Cebrecos}
\email{alcebrui@epsg.upv.es}
\author{V. Romero-Garc\'ia}
\author{R. Pic\'o}
\author{I. P\'erez-Arjona}
\author{V. Espinosa}
\author{V. J. S\'anchez-Morcillo}
 \affiliation{Instituto de Investigaci\'on para la Gesti\'on Integrada de Zonas Costeras, Universidad Polit\'ecnica de Valencia (UPV), Paranimf 1, 46730, Grao de Gandia, Spain}

\author{K. Staliunas}
 \affiliation{ICREA, Departament de F\'isica i Enginyeria Nuclear, Universitat Polit\`ecnica de Catalunya, Colom, 11, E-08222, Terrasa, Barcelona, Spain}

\date{\today}

\begin{abstract}
A theoretical and experimental study of the propagation of sound beams in- and behind three-dimensional sonic crystals at frequencies close to the band edges is presented. An efficient collimation of the beam behind the crystal is predicted and experimentally demonstrated. This effect could allow the design of sources of high spatial quality sound beams.

\end{abstract}


\maketitle

Focusing and propagation of sound beams is of fundamental importance in several branches of applied acoustics, such as tomography, acoustic microscopy and imaging or sonar communication. To achieve optimal focusing, and to maximize spatial quality of the sound beams, several mechanisms have been proposed in acoustics, like the use of acoustic lenses\cite{Karpelson06} or the design of Gaussian beam transducers.\cite{Huang06} Recently it has become apparent that the materials whose properties are modulated in space, also known as sonic crystals (SCs) in acoustics\cite{Sanchez98} or photonic crystals in optics \cite{Yablonovitch87, *John87}, can modify the spatial dispersion of propagating waves. This feature opens new possibilities to control the diffractive broadening of sound beams. In particular the beams can propagate in modulated material without diffraction (the effect also referred to as self-collimation), as predicted and demonstrated in optics \cite{Zengerle87} and in acoustics.\cite{Perez07} Self-collimation is based on the existence of flat segments of spatial dispersion curves (the curves of constant frequency in $\vec{k}$-space). More recently the three-dimensional (3D) self-collimation by SCs was experimentally demonstrated,\cite{Soliveres09} which is based on the formation of flat areas of the isofrequency surfaces.

In addition to non-diffractive propagation inside the SCs, the modification of the spatial dispersion can also produce phenomena outside the crystal such as lensing\cite{Bucay09, Luo03} and superlensing.\cite{Sukhovich09} These beam propagation effects behind the SCs are related with the negative diffraction inside the periodic structure. The character of the beam propagation behind the SC depends on the wave front of the beam acquired in the system. In particular, if the wave front of the beam acquires positive curvature (due to propagation in a material with negative, or anomalous diffraction), the beam can be focused behind the modulated medium, which enables above discussed lensing and superlensing effects.

Although the focusing of sound beams behind a 2D SCs is being intensively investigated\cite{Sanchez09} the overall picture of the beam formation and propagation is still unclear. Apart from the above mentioned phase transformation effects due to the negative diffraction of waves propagating inside of the SC, spatial (or angular) filtering effects also come into play. The negatively curved segments of dispersion lines are generally surrounded by the angular bandgaps, which are angular areas where sound cannot propagate. The latter results in a modification of the angular spectrum of the beams,\cite{Staliunas09} recently demonstrated in both optics\cite{Maigyte10} and acoustics.\cite{Pico11} These two beam formation mechanisms combine, and give rich possibilities of formation of the beams with desired spatial characteristics (angular distributions) and with desired character of focusing.

\begin{figure}
\includegraphics[width=80mm]{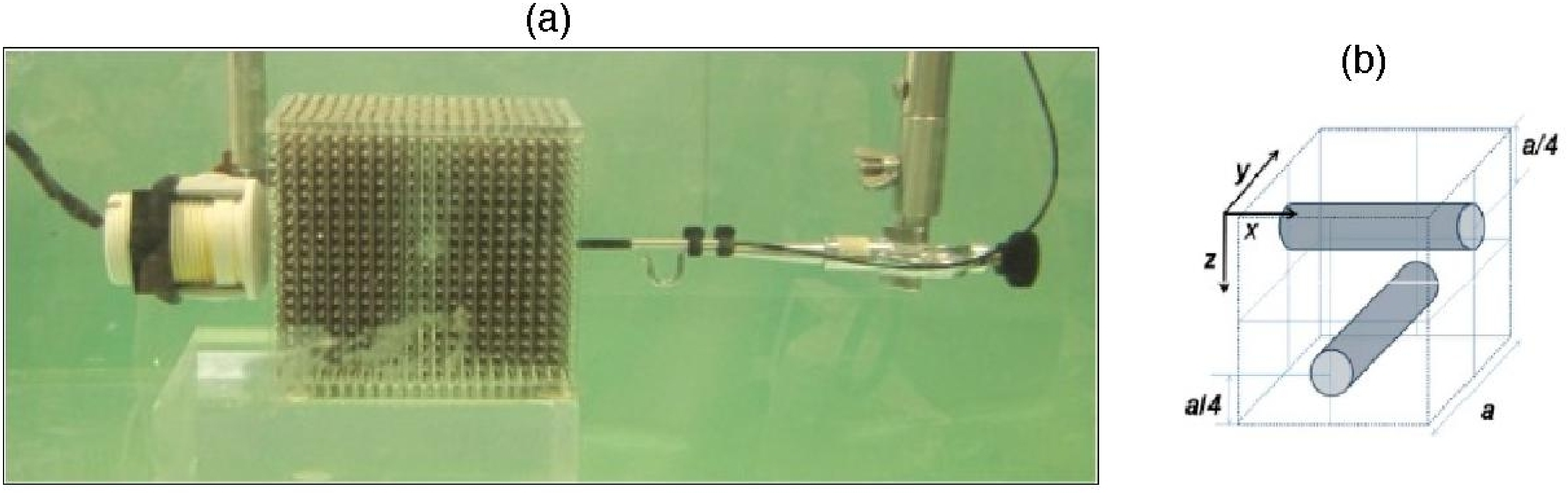}\\
\includegraphics[width=80mm]{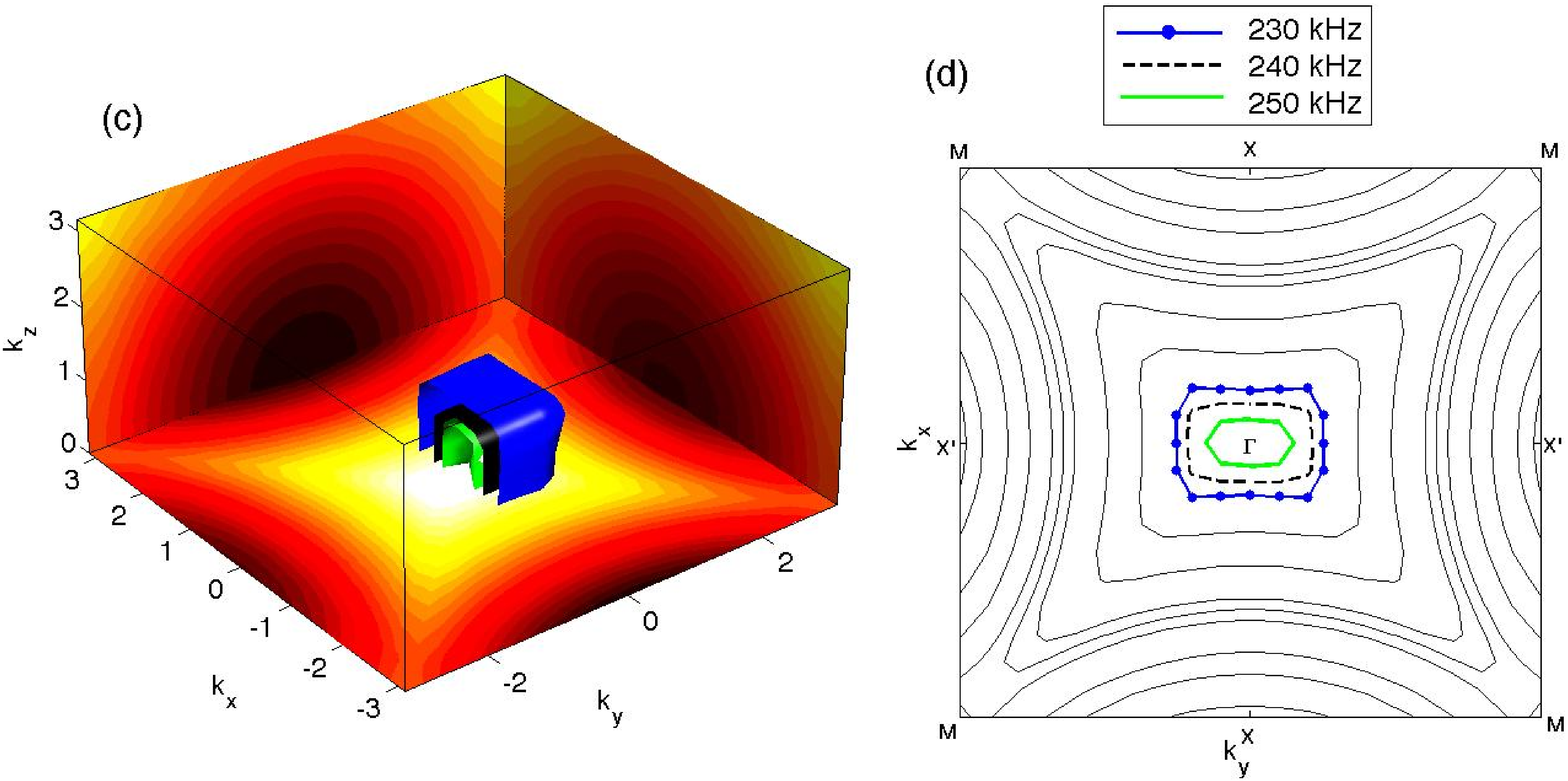}
\caption{(Color online) (a) Experimental configuration, (b) unit cell of the 3D SC, (c) isofrequency surfaces  and (d) cross-section of dispersion surfaces.}
\label{fig:Figure1}
\end{figure}

In the present work we study, experimentally and theoretically, the sound beam formation behind a 3D SC with a woodpile-like structure. We experimentally demonstrate the formation of high spatial quality and well-collimated beams behind the SC, based on the above described spatial filtering and negative diffraction effects. In this letter, we present first the isofrequency contours of SC, and identify the ranges of the frequencies where the curvature of the isofrequency surfaces is positive. Based on these results we design the samples, and perform the sound beam propagation experiments. The most important result is the experimental demonstration of the formation of a well focused beam. Finally, we investigate the beam propagation in a simplified paraxial approximation, and obtained a good quantitative interpretation of the experimental measurements. 

Figure \ref{fig:Figure1}a shows the experimental setup. The 3D SC is formed by two 2D structures of square symmetry, embedded one into another after a relative rotation by 90 degrees, which results in a 3D woodpile-like structure.\cite{Soliveres09} Each of 2D structures consist of 20$\times$20 matrix of steel cylinders of a radius $r=$0.8 mm, and the lattice constant $a=5.25$ mm (see unit cell in Fig. \ref{fig:Figure1}b). The beam, emitted by an ultrasonic source, propagates through the SC along the $z$ direction. The acoustic field is measured by a needle hydrophone positioned by a three motorized axes governed by acquisition system. As shown in Fig. \ref{fig:Figure1}a, the experimental set up is immersed in a plexiglass tank filled with distilled water.

The eigenfrequency analysis of the sound wave propagation was performed numerically using Finite Element Method.\cite{COMSOL} The periodicity of the system is considered by imposing Bloch-Floquet boundary conditions of the unit cell (Fig. \ref{fig:Figure1}b). The path around the first irreducible Brillouin zone represents the main directions of symmetry in 3D. We analyze the propagation along $\Gamma$X direction in the present work. 

Figure \ref{fig:Figure1}c shows the isofrequency surfaces for three different frequencies (230, 240 and 250 kHz) in the second band as well as the cross sections of the isofrequency surfaces by $k_z=0$, $k_x=0$ and $k_y=0$ planes respectively. Fig. \ref{fig:Figure1}c shows the quarter of the isofrequency ``bubble'' for these three frequencies. The isofrequency lines in $k_z$=0,  plane are shown in detail in Fig. \ref{fig:Figure1}d. The lowest of highlighted frequencies (230 kHz) corresponds to non-diffractive propagation inside the SC (flat isofrequency line). The isofrequency surfaces (and the lines in the cross plane) at slightly higher frequencies have areas with a positive curvature, which cause the desired focusing behaviour.

\begin{figure}
\includegraphics[width=85mm]{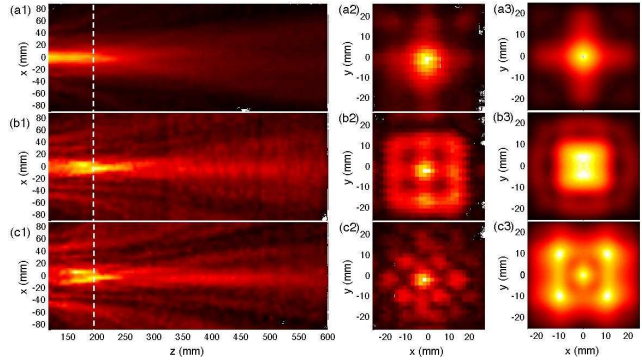}
\caption{(Color online) Experimental measurements and simulations of the acoustic field pattern behind the SC at (a) 235 kHz,  (b) 250 kHz and (c) 260 kHz. (1) XZ-cross-section of the beam behind the SC; experimental (2) and numerical (3)  ultrasound field distribution in the XY plane at point $z_1$=195 mm behind the SC.}
\label{fig:Figure2}
\end{figure}

The experimental measurements of the beams propagating behind the SC are summarized in Fig. \ref{fig:Figure2}. Three different frequencies are represented in (a) upper, (b) middle, and (c) bottom panels. The upper panel (235 KHz) shows the beam propagation for frequency corresponding to self-collimation inside the crystal.\cite{Soliveres09} The bottom panels in Fig. \ref{fig:Figure2} (260 kHz) show the beam propagation for the case when a strongly curved and relatively small ``bubble'' of isofrequency surface occurs (Fig. \ref{fig:Figure1}c). As the area of the isofrequency surface responsible for the negative diffraction and eventually for focusing is very small, just the central (paraxial) part of the angular spectrum is focalized. One part of the remaining angular components is reflected, as it corresponds to the angular bandgaps. The other part of angular components propagates along different directions, giving rise to side-lobes as seen in Fig. \ref{fig:Figure2}c. The intermediate situation, corresponding to the frequency 250 kHz, is shown in the middle panels of Fig. \ref{fig:Figure2}. The diffraction is negative for the propagation inside the SC due to positive curvature of the dispersion curves (see Figs. \ref{fig:Figure1}c and \ref{fig:Figure1}d). The isofrequency ``bubble'' is large enough to transmit a sensible portion of the angular spectrum. This case is most relevant for the goals of this work.

\begin{figure}
\includegraphics[width=85mm]{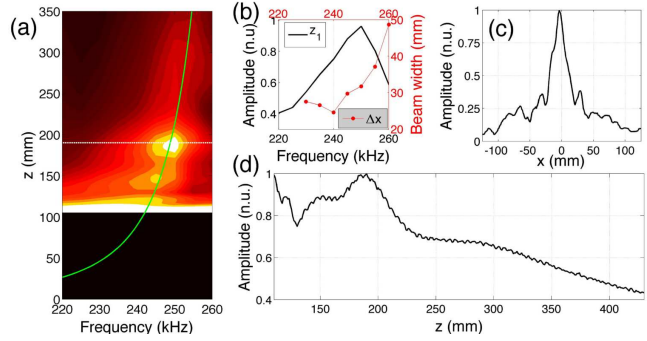}
\caption{(Color online) (a) Density plot representing the measurements of the on-axis intensity distribution on the frequency-z plane (distances measured from the transducer). Continuous green line corresponds to analytical fit (see text). Black area represents the space occupied by the SC. The white dashed line represents the point $z_1$ (b) Continuous line represents the measured beam amplitude in normalized units at point $z_1$ and red dotted line represents the measured beam width. (c) Experimental profiles in the $x$-axis of the beam at $z_1$ point and (d) measured amplitude in normalized units on the $z$-axis at 250 kHz.}
\label{fig:Figure3}
\end{figure}

We interpret the focusing of the beam in the terms of Ref. \onlinecite{Sanchez09}: the sound beam propagating in bulk of SC with negative diffraction accumulates the increasing positive (anomalous) curvature of the beam wave front. Behind the SC, the propagation in the (normally diffracting) homogeneous medium compensates the accumulated negative diffraction acquired inside the SC. The beam is focalized at some distance $z_f$ where the negative diffraction inside the SC and positive diffraction behind the SC compensate one another. 

The analytical estimation of the focal distance is possible considering the approximation of small filling fraction of the SC, $f=V_s/V_{uc}$ ($V_s$ and $V_{uc}$ are the volume occupied by the scatterer and the unit cell respectively). In this approximation the diffraction coefficient (i.e. the curvature of the spatial dispersion curve, and/or surface) can be analytically calculated.\cite{Sanchez09} Following the above interpretation, the negative diffraction of the SC is compensated at a distance $z_f$ behind the SC (measured from the input plane of the SC):   
\begin{eqnarray}
z_f=L \frac{f^2}{\Delta \Omega^3}
\label{eq:Eq1}
\end{eqnarray}
where $L$ is the length of the SC, and $\Delta\Omega=(\Omega_g-\Omega)/\Omega_g$ with $\Omega_g=\omega_ga/2\pi c_h$ being the normalized Bragg frequency and $c_h$ the speed of sound of the host medium, i.e. in water.

The experimental results were compared with the analytical study of beam focusing, as shown in Fig. \ref{fig:Figure3}. In Fig. \ref{fig:Figure3}a the absolute value of the intensity behind the crystal on the $z$-axis is mapped depending on the frequency. Green continuous line represents the analytical fit of the focal distance calculated from Eq. (\ref{eq:Eq1}) considering $L=20a$, $\Omega_g=1$, and $f=0.05$. The parameter $f$ is a fit parameter. We notice that due to the fact that Eq. \ref{eq:Eq1} has been obtained for 2D structures with low filling fraction, we use a fit parameter to take this into account. For this case, the frequency of zero diffraction point or self collimation corresponds to $\Omega_{ZDP}=(1-f^{2/3})\Omega_g$, ($\nu_{ZDP}=238$ kHz). 

The focusing for a frequency range around the optimal one is evidenced in Figs.\ref{fig:Figure3}b-d. Fig.\ref{fig:Figure3}b shows both the amplitude at point $z_1$ (black continuous line) and the beam width (red dots) depending on the frequency. Figs. \ref{fig:Figure3}c and \ref{fig:Figure3}d show the profile at $z_1$ along the $x$-axis and the transversal cross-section along the $z$ direction in $x=0$ for the focusing frequency 250 kHz respectively.

In summary, we have experimentally demonstrated the collimation of the beams behind a 3D SC. The obtained results are interpreted and analyzed in terms of curvatures of spatial dispersion curves and surfaces of the SC, and rely on the negative diffraction close to the edge of the propagating band. The experimental results fit well with the numerical simulations as well as with analytical predictions in Ref. \onlinecite{Sanchez09}. The tunability of the focal distance has been also demonstrated, showing that the beam intensity in the focus as well as the broadening of the beam along the propagation depends on the frequency, which give additional options for applications.

The overall focusing process is interpreted in terms of the interplay between two related but different effects: the focusing of the beam, due to curvature of spatial dispersion curves; and the spatial filtering effect, due to the size of the isofrequency ``bubble''. The optimum result comes from a compromise between these two ingredients.

\begin{acknowledgments}
The work was financially supported by Spanish Ministry of Science and Innovation and by European Union FEDER through projects FIS2008-06024-C02-02, -03 and MAT2009-09438. V.R.G. is grateful for the support of post-doctoral contracts of the UPV CEI-01-11. K.S. acknowledges the grant of UPV PAID-02-01.
\end{acknowledgments}

%

\end{document}